\documentclass[aps,prb,amsmath,twocolumn]{revtex4}
\usepackage{graphicx}

\DeclareMathOperator{\tr}{tr}

\def\XXint#1#2#3{{\setbox0=\hbox{$#1{#2#3}{\int}$}
\vcenter{\hbox{$#2#3$}}\kern-.5\wd0}}

\begin{document}
\title{Persistent current noise and electron-electron interactions}
\author{Andrew G. Semenov$^{1}$
and Andrei D. Zaikin$^{2,1}$
}
\affiliation{$^1$I.E.Tamm Department of Theoretical Physics, P.N.Lebedev
Physics Institute, 119991 Moscow, Russia\\
$^2$Institut f\"ur Nanotechnologie, Karlsruher Institut f\"ur Technologie (KIT),
76021 Karlsruhe, Germany
}

\begin{abstract}
We analyze fluctuations of persistent current (PC) produced by
a charged quantum particle moving in a ring and interacting with a
dissipative environment formed by diffusive electron gas.
We demonstrate that in the presence of interactions such
PC fluctuations persist down to zero temperature. In the case of
weak interactions and/or sufficiently small values of the ring radius $R$
PC noise remains coherent and can be tuned by external
magnetic flux $\Phi_x$ piercing the ring. In the opposite limit
of strong interactions and/or large
values of $R$ fluctuations in the electronic bath strongly suppress quantum
coherence of the particle down to $T=0$ and induce incoherent $\Phi_x$-independent
current noise in the ring which persists
even at $\Phi_x=0$ when the average PC is absent.

\end{abstract}

\maketitle

%\begin{multicols}

\section{Introduction}

Persistent currents (PC) in normal meso- and nanorings pierced by
external magnetic flux represent one of the fundamental consequences
of quantum coherence of electron wave functions which at sufficiently low
temperatures may persist at distances comparable with the system size.
While the average value of PC was intensively investigated
both theoretically \cite{thy} and experimentally \cite{exp} during last
decades, little was known about equilibrium fluctuations of PC.

At non-zero $T$ it is quite natural to expect non-vanishing thermal
fluctuations of PC \cite{Moskalets}. Somewhat less trivial is the limit
$T \to 0$ when the system approaches its (non-degenerate) ground state.
In this limit no PC fluctuations occur only provided the current
operator $\hat I$ commutes with the total Hamiltonian $\hat H$ of the
system, otherwise fluctuations of PC do not vanish even provided the
system remains exactly in its ground state \cite{SZ10}. For
instance, PC does fluctuate down to $T=0$ provided the ring is coupled
to some external dissipative environment.
This situation is encountered, e.g., within the model \cite{Buttiker}
where it was demonstrated that, while the average value of PC
$\langle \hat I \rangle$ in the ground state decreases with
increasing coupling with the environment, PC fluctuations remain
non-zero and even increase with the coupling constant.

In the latter example interaction with the dissipative environment
produces quantum decoherence which, in turn, causes suppression of PC
$\langle \hat I \rangle$ even at $T=0$. On the other hand, ground
state fluctuations of PC can occur also in the absence of
any decoherence. E.g., the operators $\hat I$ and $\hat H$ do
not commute for a quantum particle on a ring in the presence of
an external periodic potential. This model describes, e.g.,
the properties of superconducting nanorings with
quantum phase slips \cite{AGZ}. In this case PC fluctuations do
not vanish even in the ground state \cite{SZ10} and, at the same
time, quantum coherence remains fully preserved. As a result,
the magnitude of PC fluctuations turns out to be sensitive to
external magnetic flux \cite{SZ10}. This observation implies that
quantum coherence and decoherence in meso- and nanorings can be
probed by measuring the equilibrium current noise in such systems.
The main goal of this paper is to theoretically analyze such
current noise in the presence of quantum decoherence produced by
electron-electron interactions.

Note that the existing theory of quantum decoherence
of electrons in disordered conductors with electron-electron
interactions \cite{GZ1,GZ2} is rather complicated, merely
because of the Pauli principle which needs to be properly
accounted for in such systems. At the same time, the
basic physics of this phenomenon can be explained already without
unnecessary complications: It is due to the electron interaction with
the fluctuating quantum electromagnetic field produced by other
electrons moving in a disordered conductor.
Hence, for the sake of physical transparency it is sometimes useful to
employ a simplified model which mimics all key features of
the ``real'' problem of interacting electrons in a disordered
conductor except for the Pauli exclusion principle.
This model \cite{Paco} deals with a quantum particle
moving in a ring and interacting with some
quantum dissipative environment. The latter could be, e.g., a bath
of Caldeira-Leggett oscillators or electrons in a disordered
conductor. In the case of Caldeira-Leggett environment decoherence of
a quantum particle on a ring was investigated with the aid of
imaginary \cite{Paco} and real-time \cite{GZ98} approaches which
yield similar results, i.e. exponential suppression of quantum
coherence down to $T=0$ at sufficiently large ring radii.
In fact, the problem \cite{Paco,GZ98} is
exactly equivalent to that of Coulomb blockade in a
single electron box where exponential reduction of the effective charging energy at
large conductances is also well established \cite{pa91,HSZ}.
The model of a particle in a diffusive electron gas
was extensively used by different authors
\cite{Paco,GHZ,GSZ,HlD,CH,KH,SZ09} in order to investigate
the effect of interaction-induced decoherence on the average
value of PC. Below we will employ the same model in order
to study an interplay between PC fluctuations and quantum
decoherence.

The structure of our paper is as follows. In Sec. 2 we define our
model and describe our basic formalism of the influence
functional. In Sec. 3 we analyze PC fluctuations within the framework of
the perturbation theory in the interaction. In Sec. 4 we go beyond
the perturbation theory and evaluate equilibrium current-current
correlation functions in the limit of strong interactions.
Our main conclusions are outlined in Sec. 5. Further technical details
of our calculation are specified in Appendices A and B.

\section{The model and basic formalism}

In our analysis we will employ a model \cite{Paco,GHZ} of a quantum particle with mass $M$
on a 1d ring of radius $R$ pierced by magnetic flux $\Phi_x$. This quantum particle
interacts with a dissipative environment (bath) described by
some collective variable $V$ representing its degrees of freedom.
The total Hamiltonian for this system reads
\begin{equation}
   \hat H=\frac{(\hat \phi -\phi_x)^2}{2MR^2}+\hat H_{\rm env}(V)+\hat H_{\rm int}(\theta,V),
\label{H}
\end{equation}
where $\theta$ is the angle variable parameterizing the particle position on the ring, $\hat \phi =-i\frac{\partial}{\partial\theta}$ is the
angular momentum operator, $\phi_x=\Phi/\Phi_0$ and $\Phi_0$ is the
flux quantum. The first term in Eq. (\ref{H}) represents the particle kinetic energy, $\hat H_{\rm env}(V)$
is the Hamiltonian of the bath, and
the term $\hat H_{\rm int}(\theta,V)$ describes interaction between the particle and
the bath. Note that within our model the interaction term involves only the coordinate (not momentum) operators.

Let us define the current operator for a particle on a ring. It reads
\begin{equation}
    \hat I=\frac{e}{2\pi}\dot{\hat \theta}=\frac{ie}{2\pi}[\hat H,\hat\theta ]=\frac{e(\hat\phi -\phi_x)}{2\pi MR^2}.
\label{curop}
\end{equation}
Employing the Heisenberg representation of this operator
\begin{equation}
\hat I(t)=e^{it\hat H}\hat I e^{-it\hat H},
\end{equation}
one can define the current-current correlation function
$\langle\hat I(t)\hat I(0)\rangle$.
Evaluating the symmetrized version of this correlator in equilibrium
one arrives at PC noise power \cite{SZ10}
\begin{equation}
S(t)=\frac12\langle\hat I(t)\hat I(0)+\hat I(0)\hat I(t)
\rangle-\langle\hat I\rangle^2=\int\frac{d\omega}{2\pi}S_\omega e^{-i\omega t},
\label{sdef}
\end{equation}
which represents the central object in our subsequent analysis.

Similarly to Refs.
\onlinecite{Paco,GHZ} we will model the environment by 3d
diffusive electron gas with the inverse dielectric function
\begin{equation}
\frac{1}{\epsilon (\omega , k)}\approx \frac{-i\omega +Dk^2}{4\pi \sigma},
\label{diel}
\end{equation}
where $\sigma$ is the Drude conductivity of this gas, $D=v_Fl/3$
is the electron diffusion coefficient, $v_F$ is Fermi velocity
and $l$ is the electron elastic mean free path. As usually, below
we will assume disorder to be not very strong implying
this mean free path to be much longer than
the inverse electron Fermi momentum $k_F^{-1}$, i.e. $k_F l \gg 1$.
On the other hand, the electron mean free path should remain much
smaller than the ring radius $l \ll R$. We also note that Eq.
(\ref{diel}) applies at frequencies $\omega\ll \omega_c \sim v_F/l$.

Fluctuating electrons cause fluctuations of
the electric potential $V$ in the system. Within Gaussian approximation
such fluctuations are described by the correlator
\begin{equation}
\langle VV\rangle_{\omega , k}= -\coth \frac{\omega}{2T}{\rm Im}\frac{4\pi}{k^2\epsilon (\omega , k)}.
\end{equation}
Interaction between the particle on a ring
and fluctuating electrons in the environment is described by the
standard Coulomb term
\begin{equation}
\hat H_{\rm int}=e\hat V, \label{eV}
\end{equation}
where $e$ denotes the particle charge.

In order to evaluate the correlation function (\ref{sdef}) it is necessary to describe quantum
dynamics of our system. For this purpose let us introduce the evolution operator
$\hat U(t,t_0)$ and define the density matrix operator
\begin{equation}
  \hat \rho(t)=\hat U(t,0)\hat\rho_i \hat U^\dag(t,0),
\label{rhot}
\end{equation}
where $\rho_i$ is the initial density matrix.
The kernel of the evolution operator can be expressed as a path integral over the angle variable. We have
\begin{eqnarray}
\langle \theta_1 |\hat U(t,0)|\theta_1'\rangle = e^{i\phi_x(\theta_1-\theta_1')}\nonumber\qquad\qquad\qquad \\ \times\sum\limits_{m=-\infty}^\infty e^{2\pi i m\phi_x}U(\theta_1+2\pi m,t;\theta_1',0),
\end{eqnarray}
where
\begin{equation}
U(\theta_1,t;\theta_1',0)=\int\limits_{\theta(0)=\theta_1'}^{\theta(t)=\theta_1}  \mathcal D V\mathcal D \theta e^{iS+iS_{\rm env}},
\end{equation}
 $S_{\rm env}$ is the action of the environment and
\begin{equation}
S=\int\limits_{0}^t \left(\frac{\dot \theta^2(t_1)}{4E_C}-eV({\bf r}_\theta(t_1),t_1)\right)dt_1.
\end{equation}
Here $E_C=1/(2MR^2)$ and  ${\bf r}_\theta$ is the vector with components  $(R\cos \theta, R\sin \theta )$.

As we are interested in the dynamics of the particle rather than that of the bath, it is
convenient to trace out
fluctuating potential of the environment $V$. Making use of a standard simplifying assumption
that at the initial time moment the total density matrix is factorized into the product of the equilibrium bath density matrix and some initial particle density matrix $\hat \rho_i$, we obtain
\begin{widetext}
\begin{eqnarray}
  \rho(\theta_1,\theta_2;t)=\sum\limits_{m_1,m_2=-\infty}^\infty e^{i(\theta_1+2\pi m_1)\phi_x-i(\theta_2+2\pi m_2)\phi_x} \int\limits_0^{2\pi}d\theta_1'd\theta_2' e^{-i(\theta_1'-\theta_2')\phi_x}\rho_i(\theta_1',\theta_2')\nonumber\\
  \times\int\limits_{\theta^F(0)=\theta_1'}^{\theta^F(t)=\theta_1+2\pi m_1}\mathcal D \theta^F\int\limits_{\theta^B(0)=\theta_2'}^{\theta^B(t)=\theta_2+2\pi m_2}\mathcal D \theta^B e^{i\int\limits_{0}^{t}[((\dot \theta^F)^2-(\dot \theta^B)^2)/4E_C]dt'}\mathcal F[\theta^F,\theta^B],
\end{eqnarray}
where $\rho(\theta_1,\theta_2;t)\equiv\langle\theta_1|\hat\rho(t)|\theta_2\rangle$ and
\begin{equation}
  \mathcal F[\theta^{F},\theta^B]=\left\langle e^{-i\int\limits_{0}^{t}\left(eV^F({\bf r}_{\theta^F}(t'),t')-eV^B({\bf r}_{\theta^B}(t'),t')\right)dt' }\right\rangle_V=\exp (-iS_R-S_I),
\label{ifu}
\end{equation}
is the influence functional \cite{FH} which depends on the angle variables
on the forward and backward parts of the Keldysh contour, respectively $\theta^F$ and $\theta^B$.
Calculation of this influence functional amounts to averaging over the quantum variable $V$ which is also defined on the Keldysh contour. This procedure \cite{GZ1} can easily be adapted to our present situation of a particle on a ring where no Pauli exclusion principle needs to be taken into account, cf., e.g. \cite{GSZ}.  Introducing the new variables  $\theta_+=(\theta^F+\theta^B)/2$ and $\theta_-=\theta^F-\theta^B$, after the standard algebra (see Appendix A for further details) we obtain
\begin{equation}
S_{R}[\theta_{+},\theta_-]=\pi\alpha \sum\limits_{n=1}^\infty a_n n \int\limits_{0}^{t} dt' \dot \theta_{+}(t')\sin(n\theta_-(t')),
\label{inffunc1b}
\end{equation}
and
\begin{equation}
  S_{I}[\theta_{+},\theta_-]=-2\pi\alpha \sum\limits_{n=1}^\infty a_n\int\limits_{0}^{t} dt' \int\limits_{0}^{t} dt''\frac{\pi T^2}{\sinh^2(\pi T(t'-t''))}\cos(n(\theta_{+}(t')-\theta_{+}(t'')))\sin\frac{n\theta_-(t')}{2}\sin\frac{n\theta_-(t'')}{2},
  \label{inffunc1a}
\end{equation}
\end{widetext}
where $\alpha=3/(8 k_F^2l^2)$ is the effective coupling constant in our problem and $a_n$ are the Fourier coefficients equal to $a_n=(2/(\pi r))\ln(r/n)$ for $n<r\equiv R/l \gg 1$ and to zero $a_n=0$ otherwise. The weak disorder condition $k_Fl \gg 1$ obviously implies
$\alpha \ll 1$, i.e. the coupling constant always remains small within the applicability range of our model.

We also note that the above influence functional reduces to the Caldeira-Leggett one if we choose \cite{Paco,GHZ} $\alpha = \eta R^2/\pi$ (where $\eta$ is effective viscosity of the Caldeira-Leggett bath), $a_1= 1$ and $a_n=0$ for all $n>1$.

\section{Perturbation theory}
Let us assume that both effective coupling constant $\alpha$ and ring radius $R$ are sufficiently small
and proceed within the perturbation theory in the interaction. Keeping only the first order correction
$\sim \alpha$ to the average value of PC, in the low temperature limit $T \ll E_C$ one finds \cite{GHZ}
\begin{equation}
\langle \hat I \rangle = \frac{eE_C}{\pi}\left [\phi_x-\frac{\alpha}{2}\sum_{n=1}^{r}na_n\ln \left (
\frac{n+2\phi_x}{n-2\phi_x}\right )\right ],
\label{PCpert}
\end{equation}
for $-1/2<\phi_x <1/2$. The result (\ref{PCpert}) should be periodically continued outside this interval
of flux values. We observe that for $\alpha \ll 1$ the first order perturbative correction to the average
value of PC is negligibly small except in the immediate vicinity of half-integer flux values $\phi_x=\pm 1/2,
\pm 3/2, ...$, where the two lowest energy levels become very close to each other and the perturbation theory fails already in the first order.

Turning now to PC noise let us recall that in the limit $\alpha \to 0$ the current operator (\ref{curop}) commutes with the total Hamiltonian. Hence, in the absence of interactions PC noise vanishes in the low temperature limit for the model in question \cite{SZ10}. On the
other hand, in the presence of interactions PC noise remains non-zero and important contributions to the current-current correlator (\ref{sdef}) are expected to occur already at sufficiently small values of $\alpha$ and $R$. This limit will be studied perturbatively below in this section.

\subsection{Density matrix and Dyson equation}

\begin{figure}[t]
\includegraphics[width=0.9\columnwidth]{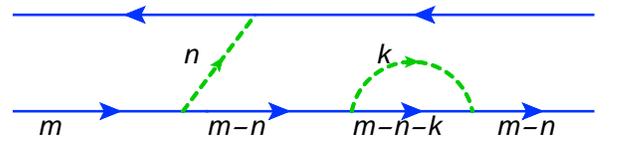}
\caption{Typical Keldysh diagram contributing to the density matrix.}
\label{typdiag}
\end{figure}

In order to proceed it will be convenient for us to pass to the momentum representation. Performing the Fourier transformation of the density matrix
\begin{equation}
\tilde \rho (m,n;t)=\int\limits_0^{2\pi}\frac{d\theta_1}{2\pi}\int\limits_0^{2\pi}\frac{d\theta_2}{2\pi}\rho (\theta_1,\theta_2;t) e^{-im\theta_1+in\theta_2}.
\end{equation}
and making use of the influence functional in the form (\ref{inffunc2}) (see Appendix A), we obtain
\begin{eqnarray}
\tilde \rho (m_1,m_2;t)=\sum\limits_{m_1',m_2'=-\infty}^\infty \tilde\rho_i (m_1',m_2')\qquad
\nonumber\\
\times\left\langle\tilde K(m_1,t;m_1',0) \tilde K^*(m_2,t;m_2',0)\right\rangle_{\nu_n},
\label{rhoex}
\end{eqnarray}
where
\begin{eqnarray}
  \tilde K(m,t;m',0)=\int\limits_{-\infty}^\infty d\theta_1\int\limits_0^{2\pi}\frac{d\theta_1'}{2\pi} e^{-i(m-\phi_x)\theta_1+i(m'-\phi_x)\theta_1'}\nonumber\\
  \times\int\limits_{\theta(0)=\theta_1'}^{\theta(t)=\theta_1}\mathcal D \theta e^{i\int\limits_{0}^{t}\left(\frac{\dot \theta^2}{4E_C}+\sum\limits_{n=1}^\infty(\nu_n(t')e^{in\theta (t')}+c.c.)\right)dt'}
\label{Ka}
\end{eqnarray}
Let us expand the exponent in Eq. (\ref{Ka}) in series in $\nu_n$ and evaluate Gaussian path integrals
 in all terms of this expansion. In the zeroth order (all $\nu_n=0$) we obtain
\begin{eqnarray}
   \tilde K(m,t;m',0)=\delta_{m,m'}K_0(m;t)
\label{zeroth}
\end{eqnarray}
with $K_0(z;t)=e^{-iE_C(z-\phi_x)^2t}$, the term with $\nu_1=1$ and $\nu_{n\neq 1}$ yields the contribution
\begin{eqnarray}
\delta_{m-n,m'} K_0(m;t-t_1) K_0(m-n;t_1),
\end{eqnarray}
and so on. Representing each term in this expansion diagrammatically, we can indicate the unperturbed propagator $\delta_{m,m'}K_0(m;t)$ (\ref{zeroth}) by a solid line and observe that each insertion of $e^{in\theta(t)}$ (or $e^{-inx(t)}$) adds (or removes) the momentum $n$ at a time $t$. Collecting all contributions to all orders in $\alpha$ in Eq. (\ref{rhoex}) and averaging over $\nu_n$ we arrive at the perturbation series for the density matrix which can be expressed in terms of Keldysh diagrams consisting of two solid lines (implying forward and backward propagators) connected by dashed lines corresponding to the propagators $\Pi^{\sigma,\sigma'}_n(t-t')$ for the $\nu_n$-field. One of such diagrams is depicted in Fig. \ref{typdiag} and is similar to Keldysh diagrams encountered, e.g., in the Coulomb blockade problem \cite{SS}.

Let us denote the sum of all diagrams contributing to the evolution kernel for the particle density matrix as $\mathcal U_{m_1,m_1'}^{m_2,m_2'}(t)$, where lower (upper) indices correspond to forward (backward) lines.  Provided upper and lower indices coincide, only lower indices will be indicated, i.e. $\mathcal U_{m,m'}^{m,m'}(t)\equiv \mathcal U_{m,m'}(t)$. The latter quantities can also be considered as
elements of the matrix  $\hat {\mathcal U}(t)$. In what follows we will also use the short-hand notation for the density matrix $\mathcal P_{m_1}^{m_2}(t) \equiv \tilde \rho (m_1,m_2;t)$ and denote its diagonal elements as $\mathcal P_m(t)$ or as a ket-vector $|\mathcal P(t)\rangle$. In this notations the evolution of the density matrix  is expressed by means of the equation
\begin{equation}
\mathcal P_{m_1}^{m_2}(t)=\sum\limits_{m_1',m_2'}\mathcal U_{m_1,m_1'}^{m_2,m_2'}(t)\mathcal P_{m_1'}^{m_2'}(0).
\end{equation}
Introducing the self-energy $\mathcal S_{m_1,m_1'}^{m_2,m_2'}(t-t')$ in a standard manner as a sum of all irreducible diagrams we arrive at the following Dyson equation for the kernel of the evolution operator
\begin{widetext}
\begin{eqnarray}
\mathcal U_{m_1,m_1'}^{m_2,m_2'}(t)=K_0(m_1;t)K_0^*(m_2;t)\delta_{m_1,m_1'}\delta_{m_2,m_2'}+\sum\limits_{n_1,n_2}\int\limits_{0}^{t}dt_1\int\limits_{t_1}^{t}dt_2\qquad\qquad \nonumber\\ \times K_0(m_1;t-t_2)K_0^*(m_2;t-t_2)\mathcal S_{m_1,n_1}^{m_2,n_2}(t_2-t_1)\mathcal U_{n_1,m_1'}^{n_2,m_2'}(t_1).
\label{dysoneq}
\end{eqnarray}
\end{widetext}

Suppose that our system was described by diagonal density matrix at some initial time which tends to $-\infty$. Then it should remain in the diagonal state at all later times as well. The evolution of such density matrix is determined by the equation
\begin{equation}
\frac{\partial|\mathcal P(t)\rangle}{\partial t}=\int\limits_{-\infty}^tdt'\hat {\mathcal S}(t-t')|\mathcal P(t')\rangle .
\label{mastereq}
\end{equation}
At long enough times $|\mathcal P(t)\rangle$ tends to its equilibrium (and, hence, time-independent) value $|\mathcal P^{eq}\rangle$ which obeys the equation
\begin{equation}
\int\limits_{-\infty}^tdt' \hat {\mathcal S}(t-t')|\mathcal P^{eq}\rangle =0.
\end{equation}
Performing the Fourier transformation of the self-energy $\hat {\mathcal S}_\omega=\int\limits_0^\infty dt e^{i\omega t}\hat {\mathcal S}(t)$ one can rewrite the above equation as $\hat {\mathcal S}_0 |\mathcal P^{eq}\rangle=0$. In equilibrium one obviously has $\mathcal P_m^{eq}=e^{-E_m/T}/\mathcal Z$, where $\mathcal Z=\sum\limits_m e^{-E_m/T}$ is the partition function and $E_m=E_C(m-\phi_x)^2$.

\subsection{Self-energy}

The evolution kernel for diagonal elements of the density matrix can also be expressed via the self-energy as
\begin{equation}
\hat{\mathcal U}_\omega=\frac{i}{\omega-i\hat{\mathcal S}_\omega}.
\end{equation}
Since all elements of the ket-vector $|\mathcal P(t)\rangle$ are real, one can demonstrate that the self-energy remains purely real in the zero frequency limit. Introducing the matrices $\hat \Sigma_\omega$ and $\hat \Gamma_\omega$ (which are non-singular at small frequencies) we can write
\begin{equation}
\hat{\mathcal S}_\omega=i\omega\hat\Sigma_\omega-\hat\Gamma_\omega
\end{equation}
and, hence,
\begin{equation}
   \hat{\mathcal U}_\omega=\frac{i}{\omega+i(1+\hat\Sigma_\omega)^{-1}\hat \Gamma_\omega }(1+\hat\Sigma_\omega)^{-1}.
\end{equation}
Here we employ a simple approximation which amounts to neglecting $\hat \Sigma_\omega$ and keeping only the leading in $\alpha$ correction to $\hat \Gamma_\omega$. Then we obtain
\begin{equation}
  \hat{\mathcal U}_\omega^{(0)}\approx\frac{i}{\omega+i\hat\Gamma_\omega^{(0)}},
\end{equation}
\begin{figure}[t]
\includegraphics[width=0.9\columnwidth]{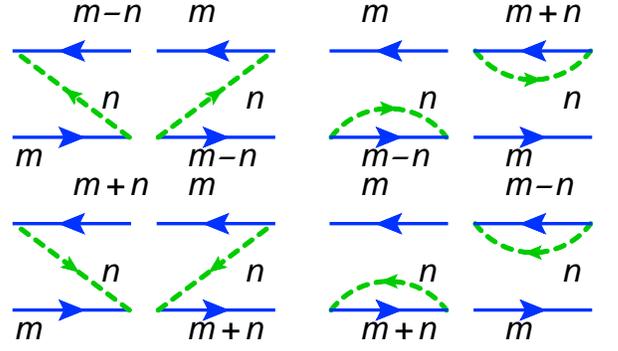}
\caption{First order self-energy diagrams}
\label{selfen}
\end{figure}
where the real part of the self-energy is defined as
\begin{eqnarray}
[\hat \Gamma_\omega^{(0)}]_{m+n,m}=-\frac{\pi\alpha a_{|n|}}{2}\left(\frac{E_{m+n}-E_m+\omega}{e^{\frac{E_{m+n}-E_m+\omega}{T}}-1}\right.\qquad\nonumber\\
\left.+\frac{E_{m+n}-E_m-\omega}{e^{\frac{E_{m+n}-E_m-\omega}{T}}-1}\right),
\end{eqnarray}
\begin{equation}
[\hat \Gamma_\omega^{(0)}]_{m,m}=-\sum\limits_{n=1}^\infty\left([\hat \Gamma_\omega^{(0)}]_{m+n,m}+ [\hat \Gamma_\omega^{(0)}]_{m-n,m}\right).
\end{equation}
The corresponding first order self-energy diagrams are depicted in Fig. \ref{selfen}.

\subsection{Current-current correlator}

\begin{figure}[t]
\includegraphics[width=0.9\columnwidth]{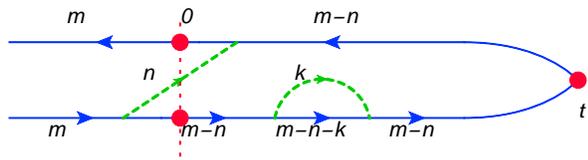}
\caption{Typical diagram contributing to the current-current correlator (\ref{sdef+}).}
\label{currcorrd}
\end{figure}

Let us identically rewrite the current noise power (\ref{sdef}) in the form
\begin{equation}
   S_\omega=2\Re\int\limits_0^\infty dt e^{i\omega t}\langle I_+(t) I_+(0)\rangle-2\pi\delta(\omega)\langle I_+\rangle^2,
\label{sdef+}
\end{equation}
where we defined $I_+=(I^F+I^B)/2$. A typical diagram contributing to this expression is depicted in Fig. \ref{currcorrd}. Let us recall that in our problem the current and momentum operators coincide with each other up to a constant. Hence, inserting the current operator at a time $t$ inside the diagram
yields the factor $eE_C(m-\phi_x)/\pi$, where $m$ is the particle momentum value at this time. In what follows we will divide all diagrams into two different classes, those with equal momenta at both upper and lower at $t=0$ and all others. Summing up all diagrams in each of these two classes we arrive at two different
contributions to the current-current correlator:
\begin{eqnarray}
   \langle I_+(t)I_+(0)\rangle= \langle E|\hat{\mathcal I}^{(0)}\hat{\mathcal U}(t)\hat{\mathcal I}^{(0)}|\mathcal P^{eq}\rangle\qquad\qquad\nonumber\\
   +\int\limits_{-\infty}^0 dt''\int\limits_0^t dt' \langle E|\hat{\mathcal I}^{(0)}\hat{\mathcal U}(t-t')\hat{\mathcal I}^{(1)}(t',t'')|\mathcal P^{eq}\rangle ,
\end{eqnarray}
where $\hat{\mathcal I}^{(0)}$ is a diagonal matrix with elements $\hat{\mathcal I}^{(0)}_{m,n}=2E_C(m-\phi_x)\delta_{m,n}$ and $\hat{\mathcal I}^{(1)}(t,t')$ is defined as a sum of all irreducible diagrams containing $t=0$. After the Fourier transformation we obtain
 \begin{eqnarray}
   S_\omega=2\Re \langle E|\hat{\mathcal I}^{(0)}\hat{\mathcal U}_\omega(\hat{\mathcal I}^{(0)}+\hat{\mathcal I}^{(1)}_\omega)|\mathcal P^{eq}\rangle\nonumber\qquad\\
   -2\pi\delta(\omega)\langle E|\hat{\mathcal I}^{(0)}|\mathcal P^{eq}\rangle^2.
\end{eqnarray}
This equation defines a formally exact expression for PC noise power. Proceeding perturbatively in $\alpha$ one
can check that the contribution containing $\hat{\mathcal I}^{(1)}_\omega$ can be neglected as
it turns out to be small as $\sim \alpha \ll 1$ as compared to the terms with $\hat{\mathcal I}^{(0)}_\omega$.
Then after some manipulations (see Appendix B for further details) we get
\begin{eqnarray}
   S_\omega=2\Re \langle E|\hat{\mathcal I}^{(0)}\left(\hat{\mathcal U}_\omega-\frac{i|\mathcal P^{eq}\rangle\langle E|}{\omega+i0}\right)\hat{\mathcal I}^{(0)}|\mathcal P^{eq}\rangle .
\label{PCnp}
\end{eqnarray}
Note that at small $\omega$ the quantity $\hat{\mathcal U}_\omega$ tends to $i|\mathcal P^{eq}\rangle\langle E|/(\omega+i0)$, i.e. PC noise power is regular in the zero frequency limit. At non-zero $\omega$
the difference between these two terms is proportional to $\alpha$, thus providing nonvanishing
PC noise at such frequencies in the presence of interactions.

\subsection{Results}

\begin{figure}[t]
\includegraphics[width=0.9\columnwidth]{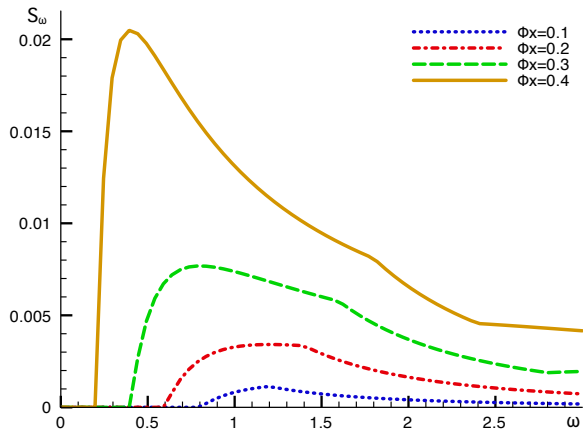}
\caption{PC noise power at $T=0$ $\pi\alpha =0.05$, $r=5$ and different flux values. Frequency $\omega$ and noise power $S_\omega$ are normalized respectively by $E_C$ and by $e^2E_C/(4\pi^2)$.}
\label{f1}
\end{figure}
\begin{figure}[t]
\includegraphics[width=0.9\columnwidth]{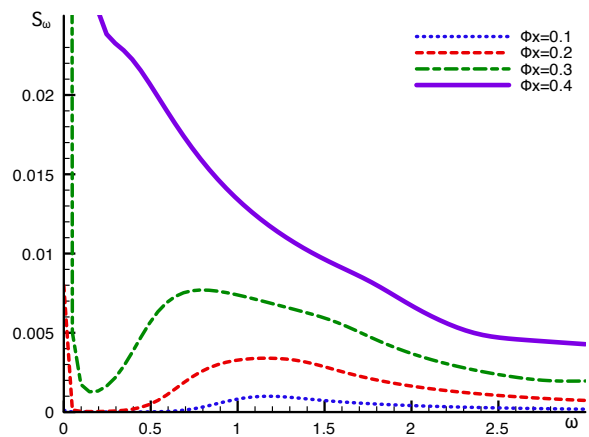}
\caption{PC noise power at $T=0.05E_C$ and different flux values. Units and parameters are the same as in Fig. \ref{f1}.}
\label{f2}
\end{figure}
\begin{figure}[t]
\includegraphics[width=0.9\columnwidth]{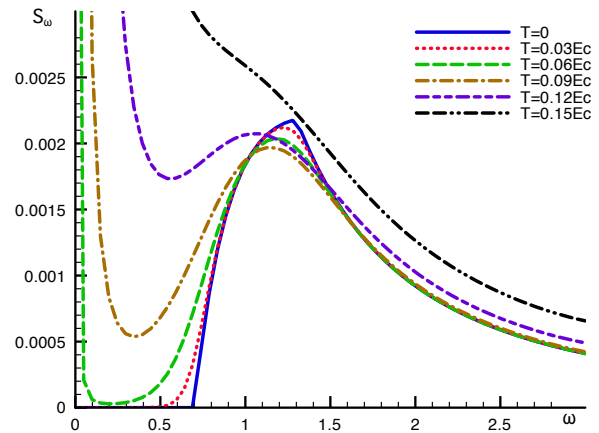}
\caption{PC noise power at $\phi_x=0.15$ and different temperatures. Units and parameters are the same as in Fig. \ref{f1}.}
\label{f3}
\end{figure}
\begin{figure}[t]
\includegraphics[width=0.9\columnwidth]{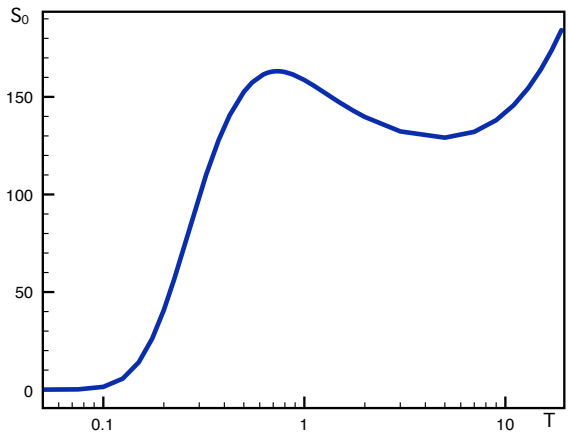}
\caption{Zero frequency PC noise power $S_0$ as a function of temperature for $\phi_x=0.15$, $\pi\alpha=0.05$ and $r=10$. Temperature and PC noise power are normalized respectively by $E_C$ and by $e^2E_C/(4\pi^2)$.}
\label{f4}
\end{figure}
The perturbative expression for the current noise power (\ref{PCnp}) was evaluated numerically at different temperatures and flux values. The results are presented in Figs. \ref{f1}-\ref{f3}. One observes that the noise power strongly depends on the magnetic flux $\phi_x$. This property illustrates the coherent nature of PC noise \cite{SZ10}. PC noise grows with increasing $\phi_x$ and diverges as the flux approaches the point $\phi_x=0.5$. This divergence has the same physical origin as that in Eq. (\ref{PCpert}). In this limit the distance between the two lowest energy levels $\delta E(\phi_x)=E_C(1-2|\phi_x|)$ becomes small and the system undergoes rapid transitions between these energy states. As these levels correspond to different PC values such transitions, in turn, yield strong current fluctuations.

It is important to observe that PC noise persists down to $T \to 0$. In this case $S_\omega$ remains zero at frequencies smaller than the inter-level distance $\omega <\delta E(\phi_x)$ and becomes non-zero otherwise. We also note that in the lowest ($\sim \alpha$) order of the perturbation theory zero temperature PC noise vanishes at $\phi_x=0$. In general, however, this feature does not hold, as it can be observed, e.g.,
from the expression for PC noise power in terms of the exact eigenstates of the total Hamiltonian \cite{SZ10}.  Non-zero PC noise at $\phi_x=0$ will also be demonstrated in the next section where we employ non-perturbative quasiclassical analysis of the problem.

At non-zero $T$ there appears additional zero frequency noise power peak. This peak grows rapidly with increasing temperature and eventually assimilates all other peaks. As a result, at sufficiently high temperatures only a wide hump remains, and PC noise becomes flux-independent, i.e incoherent.
The dependence of zero frequency PC noise on temperature is illustrated in Fig. \ref{f4}. Interestingly,
at low enough $T$ this dependence turns out to be non-monotonous, while in the high temperature limit it
approaches the linear dependence $S_0\propto T$, as it will also be demonstrated in the next section.

\section{Non-perturbative analysis}
\begin{figure}[t]
\includegraphics[width=0.9\columnwidth]{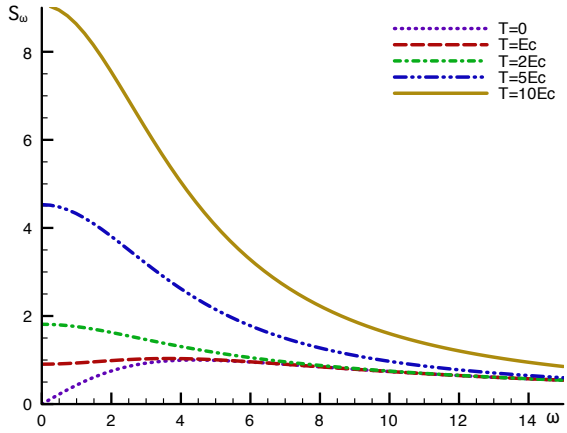}
\caption{Noise power at different temperatures for $\pi\alpha =0.05$ and $r=25$. Units 
are the same as in Fig. 4.}
\label{f5}
\end{figure}

Let us now turn to the limit of strong interactions in which case the effect of a dissipative environment on the particle motion becomes large
substantially reducing fluctuations of the angle variable $\theta$. It is important to stress that for the model under consideration this situation can be realized even at small values of the effective coupling constant $\alpha \ll 1$ provided the ring radius becomes sufficiently large \cite{GHZ}, i.e.
\begin{equation}
4\pi\alpha r \gg 1.
\label{npl}
\end{equation}
In this limit and provided temperature is not too low it suffices to employ the semiclassical approximation and to expand the effective action (\ref{inffunc1b}), (\ref{inffunc1a}) up to quadratic in $\theta_{-}$ terms. As usually \cite{Schmid,AES,GZ92}, the resulting effective action can be exactly rewritten in terms of the quasiclassical Langevin equation
for the "center-of-mass" variable $\theta_+$. For the model studied here this Langevin equation takes the form
\begin{eqnarray}
-\frac{1}{2E_C}\ddot \theta_+(t)-\frac{\gamma}{2}\dot \theta_+(t)=\sum\limits_{n=1}^\infty (\xi_n(t)\cos(n\theta_+(t))\quad\nonumber\\+\lambda_n(t)\sin(n\theta_+(t))),
\label{langev}
\end{eqnarray}
where we defined
\begin{equation}
\gamma=2\pi\alpha \sum\limits_{n=1}^\infty a_n n^2=4\pi\alpha r^2
\end{equation}
and introduced Gaussian stochastic fields $\xi_n(t)$ with the correlators
\begin{eqnarray}
\langle\xi_n(t)\xi_m(t')\rangle_{\xi,\lambda}=\langle\lambda_n(t)\lambda_m(t')\rangle_{\xi,\lambda}=\qquad\nonumber\\=-\delta_{m,n}\pi\alpha a_n n^2\frac{\pi T^2}{\sinh^2(\pi T(t-t'))},
\label{cor1}
\end{eqnarray}
\begin{equation}
\langle\xi_n(t)\lambda_m(t')\rangle_{\xi,\lambda}=0.
\label{cor2}
\end{equation}
In the high temperature limit these correlators reduce to those describing the white noise
\begin{equation}
\langle\xi_n(t)\xi_m(t')\rangle_{\xi,\lambda}=2\delta_{m,n}\pi\alpha a_n n^2T\delta(t-t')
\label{wn}
\end{equation}
and the Langevin equation can be solved exactly.  As a result, we arrive at the high temperature noise power
\begin{equation}
  S_\omega=\frac{e^2\gamma T E_C^2}{\pi^2(\omega^2+(\gamma E_C)^2)}.
\label{htpn}
\end{equation}
At lower temperatures the white noise approximation (\ref{wn}) becomes inaccurate and Eqs. (\ref{cor1}), (\ref{cor2}) should be employed. In this case noise terms in the Langevin equation can be treated perturbatively \cite{GZ92}. Keeping only the zeroth and the first order contributions one gets the solution of Eq. (\ref{langev}) in the form
\begin{equation}
\theta_+(t)=\theta_+^{(0)}+\theta_+^{(1)}(t),
\label{fior}
\end{equation}
where $\theta_+^{(0)}$ is an arbitrary (and physically irrelevant) constant and $\theta_+^{(1)}(t)$ obeys the equation
\begin{eqnarray}
-\frac{1}{2E_C}\ddot \theta_+^{(1)}(t)-\frac{\gamma}{2}\dot \theta_+^{(1)}(t)=\sum\limits_{n=1}^\infty \xi_n(t).
\label{langev1}
\end{eqnarray}
Resolving this equation one immediately arrives at the noise power in the form
\begin{equation}
S_\omega=\frac{e^2\gamma E_C^2}{2\pi^2(\omega^2+(\gamma E_C)^2)}\omega\coth\frac{\omega}{2T},
\label{noiseHT}
\end{equation}
which again reduces to Eq. (\ref{htpn}) in the high temperature limit $T \gg \omega$.
Note that for $\omega \ll \gamma E_C$ the parameter $E_C$ drops out and
the noise power becomes
\begin{equation}
    S_\omega =\frac{e^2\omega}{2\pi^2\gamma}\coth\frac{\omega}{2T},
  \end{equation}
i.e. in this case $S_\omega \propto 1/\alpha$. For $ \omega \to 0$ this expression further reduces to $S_0 \propto T/\gamma$. The noise power $S_\omega$ (\ref{noiseHT}) is also depicted in Fig. \ref{f5} at different values of $T$.

Comparing Eqs. (\ref{htpn}), (\ref{noiseHT}) with perturbative in the interaction results
for the noise power derived in the previous section we observe a striking difference between them: While in the weak interaction limit PC noise is sensitive to the externally
applied magnetic flux $\phi_x$, in the opposite limit of strong interactions the noise power $S_\omega$ turns out to be essentially independent on $\phi_x$. The latter observation implies that in the non-perturbative limit (\ref{npl}) quantum coherence of the particle is suppressed by strong interactions with the dissipative environment. This conclusion is fully consistent with earlier results \cite{GHZ} derived for PC $\langle I\rangle$ in the limit (\ref{npl}).

Note that Eq. (\ref{noiseHT}) defines only the dominating contribution to the noise power.
In addition there also exist small corrections to this result which do show the dependence
on the external flux $\phi_x$ piercing the ring. Technically, the existence of this flux-dependent terms has to do with the fact that the angle variable $\theta$ is compact (i.e. defined on a ring). On the other hand, the Langevin equation method employed above effectively "decompactifies" our problem, thus being able to capture only the $\phi_x$-independent contributions to $S_\omega$. In order to estimate the leading $\phi_x$-dependent correction to Eq. (\ref{noiseHT}) we will make use of the approach initially developed for the problem of weak Coulomb blockade in metallic quantum dots \cite{GZ96,many,Ch}. This approach establishes the relation between the density matrices and expectation values evaluated for the problems described by the same Hamiltonian but
respectively compact and non-compact variables.
With the aid of \cite{GZ96,many,Ch} for the expectation value of the current operator one finds
\begin{equation}
\langle \hat I\rangle =\frac{eE_C}{\pi}\frac{\sum_N \tr(e^{2\pi i N(\hat \phi-\phi_x)}\hat\phi\hat\rho_{np})}{\sum_N \tr(e^{2\pi i N(\hat \phi-\phi_x)}\hat\rho_{np})}.
\end{equation}
Here $\hat \rho_{np}$ stands for the reduced equilibrium density matrix for a particle described by non-compact (i.e. defined on a straight line) variable $\theta$. Analogously the noise power is given by the autocorrelation function
\begin{widetext}
\begin{equation}
S(t)=\frac{(eE_C)^2}{2\pi^2}\frac{\sum_N \tr((\hat\phi\hat\rho_{np}+\hat\rho_{np}\hat\phi)   \hat U^\dag_{np}(t,0) \hat\phi e^{2\pi i N(\hat \phi-\phi_x)} \hat U_{np}(t,0))}{\sum_N \tr(e^{2\pi i N(\hat \phi-\phi_x)}\hat\rho_{np})},
\end{equation}
\end{widetext}
where the evolution operator $\hat U_{np}(t,0)$ is again defined for a non-compact variable $\theta$. With the aid of the path integrals one can rewrite the above equations respectively as
\begin{equation}
\langle\hat I(t)\rangle=\frac{e}{2\pi}\frac{\sum_N \langle\dot\theta_+(t)e^{2\pi i N(\dot \theta_+(t)/(2E_C)-\phi_x)}\rangle}{\sum_N \langle e^{2\pi i N(\dot \theta_+(t)/(2E_C)-\phi_x)}\rangle}
\end{equation}
and
\begin{equation}
S(t)=\frac{e^2}{4\pi^2}\frac{\sum_N \langle\dot\theta_+(t)\dot\theta_+(0)e^{2\pi i N(\dot \theta_+(t)/(2E_C)-\phi_x)}\rangle}{\sum_N \langle e^{2\pi i N(\dot \theta_+(t)/(2E_C)-\phi_x)}\rangle}.
\end{equation}
In the semiclassical limit averaging in these equations is conveniently performed within the above Langevin equation technique. With the aid of Eqs. (\ref{fior}) and (\ref{langev1}) one easily finds
\begin{equation}
\langle I\rangle = \frac{i e K(0)}{2 E_C}\frac{\sum_N N e^{-2\pi i N\phi_x-\pi^2N^2 K(0)/(2E_C^2)}}{\sum_N e^{-2\pi i N\phi_x-\pi^2N^2 K(0)/(2E_C^2)}}
\label{ppcc}
\end{equation}
and
\begin{widetext}
\begin{equation}
 S(t)=\frac{e^2 K(t)}{4\pi^2}\left(1-\frac{\pi^2K(0)}{E_C^2}\frac{\sum_N N^2 e^{-2\pi i N\phi_x-\pi^2N^2 K(0)/(2E_C^2)}}{\sum_N e^{-2\pi i N\phi_x-\pi^2N^2 K(0)/(2E_C^2)}}\right),\label{ST}
\end{equation}
\end{widetext}
where we introduced the correlator $K(t)=\langle\theta_+^{(1)}(t)\theta_+^{(1)}(0)\rangle$ which Fourier transform equals to
\begin{equation}
 K_\omega\equiv\int\limits_{-\infty}^\infty dt e^{i\omega t}K(t)=\frac{2\gamma E_C^2}{\omega^2+(\gamma E_C)^2}\omega\coth\frac{\omega}{2T}.
\end{equation}
We also obtain
\begin{equation}
K(0)=4\gamma E_C^2\int\limits_0^\infty\frac{d\omega}{2\pi}\frac{\omega}{\omega^2+(\gamma E_C)^2}\coth\frac{\omega}{2T}.
\label{K0}
\end{equation}
Logarithmic divergence contained in this integral can easily be cured if we recall that
(a) our diffusive electron gas model (\ref{diel}) is applicable only at frequencies
$\omega \ll \omega_c \sim v_F/l$ (hence, the integral in Eq. (\ref{K0}) should be cut at $\omega \sim \omega_c$) and (b) our Langevin equation approach becomes insufficient in the low temperature limit where it should be supplemented by other techniques. As a result of
these considerations we may write
\begin{eqnarray}
  K(0)=C+2E_CT\qquad\qquad\qquad\qquad\qquad\qquad\nonumber \\+\frac{2\gamma E_C^2}{\pi}\left(\ln\left(\frac{\gamma E_C}{2\pi T}\right)-\psi\left(1+\frac{\gamma E_C}{2\pi T}\right)\right),
\end{eqnarray}
where $\psi(z)$ is the digamma function and the constant $C$ effectively accounts for the low temperature behavior of our system. The value of this constant can be determined if we compare the expression for PC (\ref{ppcc}) derived here with the results of the instanton analysis \cite{GHZ}.  Since in the limit (\ref{npl}) we have $K(0) \gg E_C^2$, it suffices to keep only the terms with $N=0,\pm 1$ in Eqs. (\ref{ppcc}) and (\ref{ST}). Then comparing Eq. (\ref{ppcc}) with the result \cite{GHZ} $\langle I\rangle \propto \exp (-4\pi \alpha r)$ one may identify $C$ as
\begin{equation}
C \simeq \frac{8\alpha rE_C^2}{\pi} .
\end{equation}

Finally, for the current noise power we obtain
\begin{eqnarray}
S_\omega=\frac{e^2\gamma E_C^2\omega\coth(\omega/2T)}{2\pi^2(\omega^2+(\gamma E_C)^2)}\qquad\qquad\qquad\qquad\nonumber\\\times\left(1-\frac{2\pi^2 K(0)}{E_C^2}e^{-\frac{\pi^2 K(0)}{2 E_C^2}}\cos(2\pi\phi_x)\right).
\label{noiseTot}
\end{eqnarray}
As it was already anticipated, in the limit of strong interactions (\ref{npl}) the coherent (flux-dependent) contribution in Eq. (\ref{noiseTot}) just represents a small correction
to the main incoherent term (\ref{noiseHT}). In this respect an accurate evaluation of
this small correction may even be considered as exceeding, it suffices to demonstrate that
this $\phi_x$-dependent correction remains small in the limit (\ref{npl}). We also note that  exponential dependence of PC on the ring radius $\langle I\rangle \propto \exp (-4\pi \alpha r)$ applies down to temperatures $T \sim E_C/(4\pi \alpha r)$, whereas at even lower $T \to 0$ it crosses over to a weaker (power law) dependence (cf. Figs. 1 and 2 in Ref. \onlinecite{GHZ}) but remains strongly suppressed in the limit $\alpha r \gg 1$. Accordingly, one can expect that in the same limit the flux-dependent correction to the incoherent noise term (\ref{noiseHT}) remains small down to $T=0$, though at $T \ll E_C/(4\pi \alpha r)$
it may deviate from the form (\ref{noiseTot}). Unfortunately, quantitative non-perturbative analysis of the exact zero temperature limit appears difficult since neither Langevin equation approach nor instanton analysis \cite{GHZ} can be trusted in this limit.

\section{Conclusions}

In this paper we analyzed fluctuations of persistent current produced by a charged quantum particle moving in a ring and interacting with an environment formed by 3d diffusive electron gas. Specifically, we restricted our attention to PC noise and evaluated symmetric current-current correlation function in two different limits of weak and strong interactions. Note that although within our model the effective coupling constant $\alpha$ describing Coulomb interaction between the particle and the bath always remains small, $\alpha \ll 1$, interactions can be treated perturbatively
only for sufficiently small values of the ring radius $R$, while for larger $R$  (\ref{npl}) non-perturbative analysis of interaction effects becomes unavoidable.

In the absence of interactions within our model PC fluctuates only at non-zero $T$ and
no such fluctuations could occur provided the system remains in its ground state at $T=0$.  In the presence of interactions the current operator does not anymore commute with the total Hamiltonian of the system and fluctuations of PC generally persist down to zero temperature \cite{SZ10}. In the perturbative regime of weak interactions and at sufficiently low $T$ quantum coherence of the particle remains preserved, PC noise is coherent and, hence, the noise power $S_\omega$ can be tuned by external magnetic flux $\phi_x$. In contrast, in the limit of strong interactions (\ref{npl}) fluctuations
in the electronic bath strongly suppress quantum coherence of the particle down to
$T=0$. In this case the average value of PC $\langle I\rangle$ gets strongly
suppressed as well \cite{GHZ}, while the current noise, on the contrary, does not vanish and becomes practically flux-independent. In other words, in this regime fluctuations in the environment induce incoherent background current noise in the ring which persists
even at zero flux $\phi_x=0$ when the average PC is absent $\langle I \rangle =0$.

We also point out that, while in the perturbative limit PC noise power $S_\omega$ tends to increase with the coupling constant $\alpha$, in the non-perturbative regime the dependence of $S_\omega$ on $\alpha$ becomes more complicated, cf. Eqs. (\ref{noiseHT}) and (\ref{ST}). In particular, at sufficiently low frequencies we find $S_\omega \propto 1/(\alpha R^2)$, i.e. in this regime the noise power decreases with increasing both $\alpha$
and the ring radius $R$. On the other hand, the average PC value decreases even much stronger and, hence, the ratio $S_\omega /\langle I\rangle$ {\it increases} with
increasing $\alpha$ and $R$.

Perhaps the most important result of this paper is the prediction of (i) coherent flux-dependent fluctuations of persistent current in sufficiently small rings
and (ii) incoherent flux-independent current noise in larger rings. Thus, quantum
coherence and its suppression by interactions in meso- and nanorings can be
experimentally investigated not only by detecting the average PC value (which can
happen to be very small) but also by measuring PC noise and its dependence on the external
magnetic flux. We believe it would be interesting to perform such experiments in the near future.

\appendix
\begin{widetext}
\section{Influence Functional}

Let us present the derivation of the influence functional defined by Eqs. (\ref{ifu}), (\ref{inffunc1b}) and (\ref{inffunc1a}). In order to perform Gaussian averaging over fluctuating electric potential it is
sufficient to define only the second order voltage correlators. Introducing the variables $V^{+}=(V^F+V^B)/2$ and $V^{-}=V^F-V^B$ we can express these correlators in terms of the Green functions
\begin{eqnarray}
    \langle V^{+}({\bf r}_\theta (t),t)V^{+}({\bf r}_\chi (t'),t')\rangle &=&iG^K(2R|\sin((\theta -\chi )/2)|,t-t'),
\\
    \langle V^{-}({\bf r}_\theta (t),t)V^{+}({\bf r}_\chi (t'),t')\rangle &=&iG^A(2R|\sin((\theta -\chi )/2)|,t-t'),
\\
    \langle V^{+}({\bf r}_\theta (t),t)V^{-}({\bf r}_\chi (t'),t')\rangle &=& iG^R(2R|\sin((\theta -\chi )/2)|,t-t'),
\\
    \langle V^{-}({\bf r}_\theta (t),t)V^{-}({\bf r}_\chi (t'),t')\rangle &=& 0,
\end{eqnarray}
where the last equation is a direct consequence of causality. Here $G^R$, $G^A$ and $G^K$ are respectively retarded, advanced and Keldysh Green functions related to the dielectric function $\epsilon({\bf k},\omega)$
of the environment as follows
 \begin{equation}
    G^R({\bf k},\omega)=\frac{4\pi}{{\bf k}^2\epsilon({\bf k},\omega)}\qquad\qquad G^A({\bf k},\omega)=(G^R({\bf k},\omega))^*.
 \end{equation}
 \begin{equation}
    2G^K({\bf k},\omega)=\coth\frac{\omega}{2T}\left(G^R({\bf k},\omega)-G^A({\bf k},\omega)\right).
 \end{equation}
%\begin{equation}
 %   G^R({\bf k},\omega)=-\frac{i\omega}{\sigma {\bf k}^2}+\frac{D}{\sigma} \qquad\qquad\qquad
  %  G^K({\bf k},\omega)=-\frac{i\omega}{\sigma {\bf k}^2}\coth{\frac{\omega}{2T}}.
%\end{equation}
Combining the above expressions with Eq. (\ref{diel}), in the case of a diffusive metal we obtain
\begin{equation}
    G^R(X,t-t')=\frac{2\pi\alpha}{e^2\sqrt{(X/l)^2+1}}\delta'(t-t') \qquad\qquad
    G^K(X,t-t')=\frac{2\pi\alpha}{e^2\sqrt{(X/l)^2+1}}\frac{i\pi T^2}{\sinh^2(\pi T(t-t'))} .
\end{equation}
Gaussian averaging over the $V$-fields can now easily be performed, cf., e.g., \cite{GZ1}. As a result we arrive at Eq. (\ref{ifu}). The expression for the imaginary part of the action in this equation reads
\begin{eqnarray}
  S_{I}[\theta_{+}(t),\theta_-(t)]=-\pi\alpha\int\limits_{0}^{t} dt' \int\limits_{0}^{t} dt''\frac{\pi T^2}{\sinh^2(\pi T(t'-t''))}\left(\zeta(\theta^F(t')-\theta^F(t''))+\right.\qquad\qquad\nonumber\\\left.+\zeta(\theta^B(t')-
  \theta^B(t''))
  -\zeta(\theta^F(t')-\theta^B(t''))-\zeta(\theta^B(t')-\theta^F(t''))\right),
\end{eqnarray}
where $\zeta(x)=[4(R/l)^2\sin^2(x/2)+1]^{-1/2}$. Expanding $\zeta(x)$ in the Fourier series
\begin{equation}
 \zeta(x)=a_0-\sum\limits_{n=1}^\infty a_n\sin^2\left(\frac{nx}{2}\right)=c+\frac12 \sum\limits_{n=1}^\infty a_n\cos (nx)
\label{Fourier}
\end{equation}
with $a_n=(2/(\pi r))\ln(r/n)$ for $n<r$ and $a_n=0$ for $n>r$ and using the identity
\begin{equation}
\cos(\theta_1^F-\theta_2^F)+\cos(\theta_1^B-\theta_2^B)-\cos(\theta_1^F-\theta_2^B)-\cos(\theta_1^B-\theta_2^F)
=4\cos(\theta_{1+}-\theta_{2+})\sin\frac{\theta_{1-}}{2}\sin\frac{\theta_{2-}}{2}
\end{equation}
one arrives at  Eq. (\ref{inffunc1a}). Eq. (\ref{inffunc1b}) is recovered in a similar manner.
Let us also note that the Caldeira-Leggett environment is described by the function
$\zeta(x)=-2\sin^2(x/2)=\cos(x)-1$ which should be employed in that case instead of Eq. (\ref{Fourier}).

Let us also rewrite our influence functional in a somewhat different form,
which can be conveniently used in our perturbative calculations. Employing the definition
$(-1)^F=1$ and $(-1)^B=-1$ we obtain
\begin{equation}
\mathcal F[\theta^F,\theta^B]= \left\langle e^{ i\sum\limits_{\sigma=F,B}(-1)^\sigma\sum\limits_{n=1}^\infty\int\limits_{0}^{t}(\nu_n^{\sigma}(t')
e^{in\theta^\sigma (t')}+\nu_n^{\sigma*}(t')e^{-in\theta^\sigma (t')})dt'}   \right\rangle_{\nu_n}
\label{inffunc2}
\end{equation}
Here $\nu_n$ is Gaussian stochastic complex variable described by the correlator
\begin{equation}
\langle\nu_n^{\sigma*}(t)\nu_m^{\sigma'}(t')\rangle=\frac{i\delta_{mn}}2\left(D^K_m(t-t')+\frac{(-1)^\sigma}{2}D^R_m(t'-t)+\frac{(-1)^{\sigma'}}{2}D^R_m(t-t')\right).
\end{equation}
where
\begin{equation}
D^R_n(t-t')=\pi\alpha a_n \delta'(t-t'), \quad D^K_n(t-t')=-i\pi\alpha a_n \frac{\pi T^2}{\sinh^2(\pi T(t-t'))}.
\end{equation}
Then we obtain
\begin{equation}
\mathcal F[\theta^F,\theta^B]= e^{\sum\limits_{n=1}^\infty \sum\limits_{\sigma,\sigma'=F,B}\int\limits_{0}^\infty dt\int\limits_{0}^\infty dt' \Pi_n^{\sigma,\sigma'}(t-t')\cos(n(\theta^\sigma (t)-\theta^{\sigma'}(t')))  }
\end{equation}
with $\Pi^{\sigma,\sigma'}_{n}(t-t')=-(-1)^{\sigma+\sigma'}\langle\nu_n^{\sigma*}(t)\nu_n^{\sigma'}(t')\rangle$.
\end{widetext}

\section{Operations with singular matrices}
The evolution kernel for the diagonal density matrix $\hat{\mathcal U}_\omega$ can be expressed via the self-energy by means of the following equation
\begin{equation}
\hat{\mathcal U}_\omega=\frac{i}{\omega-i\hat{\mathcal S}_\omega}.
\end{equation}
From the identity $\langle E |\hat{\mathcal S}_\omega=0$ we conclude that the matrix  $\hat{\mathcal S}_\omega$ has zero eigenvalue with the left eigenvector $\langle E|$. Hence, there also exists the right eigenvector $|\omega\rangle$ with the same (i.e. zero) eigenvalue, $\hat{\mathcal S}_\omega|\omega\rangle=0$. Employing the normalization condition $\langle E|\omega\rangle=1$ and introducing the projector $\hat{\mathcal L}_\omega=|\omega\rangle\langle E|$  we can verify the identity
\begin{equation}
\hat{\mathcal U}_\omega=\left(1+\frac{i\xi\hat{\mathcal L}_\omega}{\omega+i0}\right)\frac{i}{\omega-i\hat{\mathcal S}_\omega+i\xi\hat{\mathcal L}_\omega},
\end{equation}
which holds for any value $\xi$. This identity implies that the matrix $\hat{\mathcal U}_\omega$ is singular at small frequencies, $\hat{\mathcal U}_\omega\propto\frac{1}{\omega+i0}$. Indeed, expanding the above expression at small frequencies we obtain
\begin{equation}
 \hat{\mathcal U}_\omega\approx\frac{i\hat{\mathcal L}_0}{\omega+i0}+\frac{1}{\xi\hat{\mathcal L}_0-\hat{\mathcal S}_0}\left(1-(1-i\hat{\mathcal S}'_0)\hat{\mathcal L}_0\right).
\end{equation}
Observing that at zero frequency the vector $|\omega\rangle$ just coincides with the equilibrium distribution function, i.e. $|0\rangle=|\mathcal P^{eq}\rangle$ and making use of equations $\hat{\mathcal S}_0=-\hat\Gamma_0$ and $\hat{\mathcal S}'_0=i\hat\Sigma_0$ we get
\begin{eqnarray}
 \hat{\mathcal U}_\omega\approx\frac{i|\mathcal P^{eq}\rangle\langle E|}{\omega+i0} \qquad \quad \qquad \qquad \qquad \qquad \qquad\nonumber\\+\frac{1}{\xi|\mathcal P^{eq}\rangle\langle E|+\hat\Gamma_0}\left(1-(1+\hat\Sigma_0)|\mathcal P^{eq}\rangle\langle E|\right)
\label{B100}
\end{eqnarray}
for $\omega\to 0$ and any value of $\xi$.

%\end{multicols}


\begin{references}
\bibitem{thy} M. B\"uttiker, Y. Imry, and R. Landauer, Phys. Lett. A {\bf 96}, 365 (1985);
H.-F. Cheung, E.K. Riedel, and Y. Gefen, Phys. Rev. Lett. {\bf 62}, 587 (1989); V. Ambegaokar and U. Eckern, Phys. Rev. Lett. {\bf 65}, 381 (1990); A. Schmid, Phys. Rev. Lett. {\bf 66}, 80 (1991); F. von Oppen and E.K. Riedel, Phys. Rev. Lett. {\bf 66}, 84 (1991); B.L. Altshuler, Y. Gefen, and Y. Imry, Phys. Rev. Lett. {\bf 66}, 88 (1991).
\bibitem{exp} L.P. Levy, G. Dolan, J. Dunsmuir, and H. Bouchiat, Phys. Rev. Lett. {\bf 64}, 2074 (1990); V. Chandrasekhar, R.A. Webb, M.J. Brady, M.B. Ketchen, W.J. Gallagher, and A. Kleinsasser, Phys. Rev. Lett. {\bf 67}, 3578 (1991); E.M.Q. Jariwala, P. Mohanty, M.B. Ketchen, and R.A. Webb  Phys. Rev. Lett. {\bf 86}, 1592 (2001); A.C. Bleszynski-Jayich, W.E. Shanks, B. Peaudecerf, E. Ginossar, F. von Oppen,
     L. Glazman, and J.G.E. Harris, Science {\bf 326}, 272 (2009).
\bibitem{Moskalets} M.V. Moskalets, Physica B {\bf 301}, 286 (2001).
\bibitem{SZ10} A.G. Semenov and A.D. Zaikin,  J. Phys.: Condens. Matter {\bf 22}, 485302 (2010).
\bibitem{Buttiker} P. Cedraschi, V.V. Ponomarenko, and M. B\"uttiker,
Phys. Rev. Lett. {\bf 84}, 346 (2000); Ann. Phys. {\bf 289}, 1 (2001).
\bibitem{AGZ} K.Yu. Arutyunov, D.S. Golubev, and A.D. Zaikin,
Phys. Rep. {\bf 464}, 1 (2008).
\bibitem{GZ1} D.S. Golubev and A.D. Zaikin, Phys. Rev. Lett. {\bf 81}, 1074 (1998); Phys. Rev. B {\bf 59}, 9195 (1999); Phys. Rev. B {\bf 62}, 14061 (2000); J. Low. Temp. Phys. {\bf 132}, 11 (2003).
\bibitem{GZ2} D.S. Golubev and A.D. Zaikin, New J. Phys. {\bf 10}, 063027 (2008); Physica E {\bf 40}, 32 (2007).
\bibitem{Paco} F. Guinea, Phys. Rev. B {\bf 65}, 205317 (2002).
\bibitem{GZ98} D.S. Golubev and A.D. Zaikin, Physica B {\bf 255}, 164 (1998).
\bibitem{pa91} S.V. Panyukov and A.D. Zaikin, Phys. Rev. Lett. {\bf 67}, 3168
(1991); J. Low Temp. Phys. {\bf 73}, 1 (1988).
\bibitem{HSZ} C.P. Herrero, G. Sch\"on, and A.D. Zaikin, Phys. Rev. B
{\bf 59}, 5728 (1999) and further references therein.
\bibitem{GHZ} D.S. Golubev, C.P. Herrero, and A.D. Zaikin, Europhys. Lett.
{\bf 63}, 426 (2003).
\bibitem{GSZ} D.S. Golubev, G. Sch\"on, and A.D. Zaikin, J. Phys. Soc. Jap. {\bf 72}, Suppl. A, 30 (2003).
\bibitem{HlD} B. Horovitz and P. Le Doussal, Phys. Rev. B {\bf 74}, 073104 (2006); {\bf 82}, 155127 (2010).
\bibitem{CH} D. Cohen and B. Horovitz, J. Phys. A: Math. Theor. {\bf 40}, 12281 (2007);
  Europhys. Lett. {\bf 81}, 30001 (2008).
\bibitem{KH} V. Kagalovsky and B. Horovitz, Phys. Rev. B {\bf 78}, 125322 (2008).
\bibitem{SZ09} A.G. Semenov and A.D. Zaikin, Phys. Rev. B {\bf 80}, 155312 (2009).
\bibitem{FH} R.P. Feynman and A.R. Hibbs, {\it Quantum Mechanics and Path
Integrals} (McGraw Hill, NY, 1965).
\bibitem{GZ94} D.S. Golubev and A.D. Zaikin, Phys. Rev. B {\bf 50}, 8736 (1994).
\bibitem{SS} H. Schoeller and G. Sch\"on, Phys. Rev. B {\bf 50}, 18436 (1994).
\bibitem{Schmid} A. Schmid, J. Low Temp. Phys. {\bf 49}, 609 (1982).
\bibitem{AES} U. Eckern, G. Sch\"on, and V. Ambegaokar, Phys. Rev. B {\bf 30}, 6419 (1984).
\bibitem{GZ92} D.S. Golubev and A.D. Zaikin, Phys. Rev. B {\bf 46}, 10903 (1992); Phys. Rev. Lett. {\bf 86}, 4887 (2001).
\bibitem{GZ96} D.S. Golubev and A.D. Zaikin, JETP Lett. {\bf 63}, 1007 (1996).
\bibitem{many} D.S. Golubev, J. K\"onig, H. Schoeller, G. Sch\"on, and A.D. Zaikin,
Phys. Rev. B {\bf 56}, 15782 (1997).
\bibitem{Ch} D. Chouvaev, L.S. Kuzmin, D.S. Golubev, and A.D. Zaikin, Phys. Rev. B {\bf 59}, 10599 (1999).
\end{references}
\end{document}